\documentclass[epsfig,12pt]{article}
\usepackage{graphicx}
\usepackage{amsmath}

\input epsf.sty
\input{tcilatex}
\makeatletter \@addtoreset{equation}{section}

\def\be{\begin{equation}}
\def\ee{\end{equation}}
\def\bea{\begin{eqnarray}}
\def\eea{\end{eqnarray}}

\newcommand{\nc}{\newcommand}
\nc{\al}{\alpha} \nc{\bib}{\bibitem} \nc{\la}{\lambda}
\nc{\C}{\mbox{\hspace{1.24mm}\rule{0.2mm}{2.5mm}\hspace{-2.7mm}
C}} \nc{\R}{\mbox{\hspace{.04mm}\rule{0.2mm}{2.8mm}\hspace{-1.5mm}
R}}
\begin{document}\title{%
\rightline{\mbox {\normalsize
{}}\bigskip}\textbf{  On  the  Superstring Realization  of  the  Yang Monopole   }}
\author{Adil  Belhaj\thanks{belhaj@unizar.es}, Pablo  Diaz\thanks{pdiaz@unizar.es}, 
Antonio  Segui\thanks{segui@unizar.es}\\
{\small  Departamento de Fisica Teorica, Universidad de Zaragoza, 50009-Zaragoza, Spain.}
} \maketitle

\begin{abstract}
\bigskip
 Based on  the result of  string/string duality,  we    construct the six dimensional  Yang monopole in 
 terms of Type IIA  wrapped  D-branes. In  particular, 
we show  that all the information of  such a  magnetic solution  can be encoded in the K3 surface  compactification  in the presence of D2  and D4-branes wrapping its non trivial cycles.  We   give   a  geometrical and  physical interpretations for the $\{+1,-1\}$   Yang monopole charges. 
 Lifting to  eleven dimensions, we relate this  Type IIA configuration  with the  heterotic M-theory  one,  given in hep-th/0607193. 
 The nature  of  the black Yang monopole is also discussed.

\textbf{Keywords}: Yang Monopole, Supertring theory,  Dualities,  M-theory,   K3 surface and Black Hole.
\end{abstract}


\newpage

\newpage
\section{Introduction}
The Yang monopole \cite{Y} was constructed as a generalization of the Dirac monopole \cite{D}. 
The  latter  measures the flux of the magnetic two-form field $F$, charged under the  $U(1)$  gauge group, 
integrated over a two-sphere that encloses the origin. The flux is quantized and the monopole can have an arbitrary
 integer magnetic charge. The charge is given by integrating the first Chern class of the gauge bundle $F/2 \pi$,  over
 the  two-sphere $S^2$. This integer corresponds to the different ways the $U(1)$ connection can wrap the circle that
 constitutes the equator of the two-sphere. The origin is singular and near it an ultraviolet divergence occurs 
 for the energy of the field configuration. 

The Yang monopole is characterized by the flux of the four-form field $Tr F \wedge F$, charged under
 the  $SU(2)$   gauge group, across the four dimensional sphere that covers the origin in a $5+1$ dimensional
 space-time. It corresponds to the conformal mapping onto $S^4$ of the BPST \cite{BPST} Euclidean
 instanton solution.  Again,  the origin  is singular but now the energy of this solitonic configuration 
is well behaved in the UV regime, although IR  divergences linearly appear. The total energy inside a four sphere
  is proportional to  its  radius.  Also the flux is quantized but now, the magnetic charge of the Yang monopole
 can take only  two values $\{+1,-1\}$ \cite{Y}. This charge, which may    correspond to  the self-dual and anti-self-dual BPST  instanton 
 configurations respectively, is given by the integral over $S^4$ of the second chern class $Tr(F \wedge F) /8 \pi^2$.

The Yang monopole can be easily generalized to higher  even dimensional space-time  in the following way \cite{HP,Tc}.
 In $2n+2$ dimensions   we take the trace of the $2n$-form $F^n$, charged under the $SO(2n)$  gauge group. Then 
we integrate the flux of this form on the $2n$-sphere surrounding a singular point where the generalized 
monopole  is located. Explicit solutions can be systematically obtained \cite{GT}.

Recently,  it has been suggested the possibility of considering the Yang Monopole as placed at the end of a string with  an  energy given by the product of the tension and the length of the string \cite {P}. This fact has opened a new line for studying  the Yang monopole in string theory.  In this regard, a  heterotic M-theory
 realization   has been proposed in  \cite{BGT}. Recall  that  M-theory is an eleven dimensional theory
 containing M2-branes and its  magnetic dual M5-branes. It  also contains  M9-branes supporting, each one, an
$E_8$ gauge  bundle   of the heterotic superstring in ten dimensions\footnote{Although for not use in our discussion, KK monopoles are involved,  which
  can be identified after   circle compactification    with  D6-branes in type IIA
 superstring theory.}\cite{HoW}.
 In this  way,  $E_8 \times  E_8$  heterotic superstring can be interpreted
  in  M-theory as a M2-brane suspended between two M9-branes \cite{St,T}. 
 It has been shown \cite{BGT} that  M5-brane may have boundaries 
on M9-branes, where the  boundary is  a 4-brane  with  an  infinite tension so its centre
 of mass is not free  to move.  This  boundary  has been identified  with  the  Yang monopole, 
which  is a singular  configuration
 of the $SU(2)$   Yang Mills gauge theory  in six dimensions.
 In the  heterotic M-theory   picture, there are  two Yang monopoles. They correspond  to the ends of 
the oriented M5-brane which stretches between two M9-branes. Each monopole (each end) is charged under an $SU(2)$ subgroup  of  $E_8$  with the  topological charges  $\{ +1,-1\}$ respectively.  On the other hand, a matrix model of the Yang  monopole has been given in  \cite{CIK}.

According to  brane physics,  $SU(2)$  Yang Mills  in  six dimensions   can  be embedded into   Type II superstring theory  in, at least,  
two  different realizations  which    are connected by the   T-duality  transformations \cite{OV,K,S,HW}.
For instance,    $SU(2)$  gauge symmetry  in Type IIB superstring   can  be identified with the gauge theory living in the word-volume
 of two  coincide  D5-branes moving on flat spaces.  Alternatively,  $SU(2)$  gauge group   can be  obtained from  Type IIA superstring 
  compactified  on  a  singular K3 surface in the presence of  D2-branes wrapped  around the   collapsing  2-cycles. 
This  way  of constructing  a gauge theory from the  singular limit of  a  Calabi-Yau compactification is  known  as  the geometric engineering method \cite{KKV}. At this
level, one might naturally ask the following questions. Is there any  realization of the Yang monopole   in terms of Type II D-branes? 
 and, in this case, what will be the  relevant geometry which can explain  its  properties? 

In this  paper, we address these questions using  the duality  between 
 Type IIA superstring    compactified on   the  K3 surface  and    heterotic superstring  on  $T^4$. In particular, 
 we  will give a  Type IIA geometric  realization of the  Yang monopole in  six dimensions.   First,  we  get the $SU(2)$
 gauge  group  as an enhanced gauge symmetry  corresponding to singular limit of the  K3 surface, where the singularity arises from
 shrinking 2-cycles inside  K3.  Then we will show 
that the  Yang monopole   can  come  up as D-branes wrapping  the K3 non-trivial cycles,  in such a way that the   above properties   are  encoded in  the K3 surface features. We  also give  a K3  dual interpretation  for  the result   reported  in  \cite{BGT}.

\section{ Type IIA  superstring   construction  of the Yang monopole}
Our starting setup  is  Type IIA  superstring on  a  K3  surface. This compactification  gives
 a six dimensions model  and  reduces the number of supercharges from thirty two to  sixteen.  The  total moduli space of this theory
 is \begin{equation}
 { \cal M}= { \cal M}^{stringy}\times {\bf R}^+=\frac{SO(20,4)}{SO(20)\times SO(4)}\times {\bf R}^+, \label{stringy}
\end{equation}
 where    ${\bf R}^+  $ describes  the choice of  the superstring coupling which is a real parameter (the dilaton). 
 ${ \cal M}^{stringy}$  parameterizes   the geometric  moduli space of the $K3$  surface and 
the values of the NS-NS $B$  fields on $K3$ \cite{V}.  Note that the R-R fields do not enter in
 this  moduli space because the Betti numbers  $b_1$ and $b_3$ of $K3$ are zero.  The same physical moduli space
 appears naturally when   heterotic  superstring   is compactified  on $T^4$. 
The origin of the first factor in (\ref{stringy})  comes from the values  of the  metric, $B$ field and  non abelian  gauge vector over $T^4$ and the second factor is the dilaton field. More details  on the six dimensional duality   can be found  in \cite{V}. 

 Based on  this duality and the  result of \cite{BGT},  we show  that the Yang monopole  in six dimensions 
   can  be embedded in Type IIA superstring
 compactified on  a  $K3$  surface  in the presence of  wrapped  D-branes.     For this purpose, let us  take 
 a local description of the $K3$ surface   where the manifold develops a $su(2)$ singularity (known as  $A_1$ singularity).  This singularity  
corresponds to a  vanishing   two-sphere.   Near  such a singular point,   the K3 surface    can be identified
 with   the asymptotically  locally  Euclidean   (ALE)  space  which   is algebraically given by
\begin{equation}
 f(x,y,z)=xy-z^2=0, \label{A1}
\end{equation}
 where $(x,y,z)$ are complex variables. 
This geometry is  singular  at $x=y=z=0$  since   it is  the only  solution   of  $f=df=0$.  It   has a nice physical 
representation as  the  target space of   the two-dimensional $N = 2$  linear sigma model with only one\footnote{For the ALE geometry $A_n$ the gauge group is $U(1)^n$.} $U(1)$ gauge symmetry  and  three chiral fields $\phi_i$, $(i=1,2,3)$  with  vector   charge  ${\vec q} =(1,-2, 1)$. This vector   satisfies  the local  Calabi-Yau condition
\begin{equation}
\sum_iq_i=1-2+1=0.
\end{equation}
The coordinates of the $ALE$   space  in (\ref{A1}) can then
be expressed in terms of  the following $U(1)$  gauge invariants
\begin{equation}
x = \phi_1^2 \phi_2,\qquad
y = \phi_3^2 \phi_2,\qquad
z = \phi_1 \phi_2\phi_3.
\end{equation}
 Equation (\ref{A1}) is related to the  D-term in the bosonic potential $V(\phi_1, \phi_2, \phi_3)$  in supersymmetric
theories with four supercharges:
\begin{equation}
V(\phi_1, \phi_2, \phi_3) = ( |\phi_1|^2- 2|\phi_2|^2+|\phi_3|^2)-R)^2, 
\end{equation}
  where  $R$  is the $U(1)$  Fayet-Iliopoulos (FI) parameter\footnote{In this way, 
one sees that the $U(1)$  Cartan subgroup of the $SU(2)$  symmetry of the singularity of $K3$  carries the gauge symmetry
of the $N = 2$  supersymmetric linear sigma model.}.  The presence of this  FI term resolves
the singularity of the potential $V(\phi_1, \phi_2, \phi_3)$.  Geometrically, this corresponds to replacing the
singular point $x = y = z = 0 $ by  the 2-sphere defined by 
\begin{equation}
 S^2:\qquad |\phi_1|^2+|\phi_3|^2=R, 
\end{equation}
 which  is  the  only  non-trivial 2-cycle   on which we can wrap D2-branes. Note that  $\phi_2$   defines the non compact  direction of the 
$A_1$  $ALE$  space. The theory is   ten dimensional Type  IIA superstring  on a K3 surface with an  infinite volume.  To get the SU(2) gauge symmetry
 only the local piece containing  the  $S^2$  is nedded. Now, the system consists of    Type IIA  D2-branes wrapping around     $S^2$.  This   gives 
 a pair of massive  vectors $W^{\pm}$,  one for each of 
the two  possible   ways   of the wrapping.
  The  masses of these particles are proportional to  the volume of  the  2-sphere. 
   They are charged under  the  $U(1)$  gauge  field obtained by decomposing the type IIA 
 three form in terms of the  harmonic form  on the  2-sphere  and the
  one form gauge field   in the $K3$   transverse  six dimensional space-time. In the limit where  the 2-sphere shrinks, 
the $W^{\pm}$ particles become  massless and, together with the one form gauge field,   generate  the $SU(2)$  adjoint representation. 
 A geometric realization of $SU(2)$  gauge symmetry in  six  dimensions is then obtained \cite{KKV}. This  will be  identified with 
the  gauge symmetry of our Yang monopole.

We have obtained the electrically charged sector, associated to D2-branes wrapping  2-cycles in 
  a K3 surface. Lifting  consistently to 11 dimensions,  the M2-brane  is encounted.  It is 
responsible of the electric particles in the compactifcation of  M-theory\footnote{ 
The scenario of type IIA superstring theory in six dimensions is lifted to seven dimensions.
 The D2-branes are replaced by M2-branes, and we obtain an  $N = 2$  gauge theory in seven dimensions.}.
The dual  M5-brane, when    reduced to  ten  dimensions,   gives rise to the D4-brane  which  is the
  magnetically charged  object in Type IIA superstring.
 The  D4-brane is in essence,   the responsible for the magnetically charged sector in   six dimensions
  as  shown   below. 
Based on  the result of  M-theory/superstring dualities in seven and six dimensions, the  magnetic Yang 
monopole can be identified  with  D4-branes,  totally  wrapped   on  the K3 surface.  As consequence, they  generate the magnetic
 objects in   the six dimensional space-time. This is expected from the fact that the D4-brane
 is the only magnetic object in  Type IIA superstring theory which  can be obtained from   the M5-brane  and gives a
zero dimensional particle  after wrapping the K3 surface.   In this way, we conjecture that  all  Yang monopole  properties should 
be derived from the K3 surface  data. Since the gauge group  origin  is linked to  the singular  limit of the geometry,  we expect that
the magnetic properties can  also be  encoded in the K3 surface.

 We will show that the
charges $\{+1,-1\}$    can have different    compatible  K3 surface interpretations.
First, the different ways in which D4-branes  are wrapped on  K3 surfaces are classified by the fourth homotopy group of K3. As seen before, in order to construct the $SU(2)$ gauge group, it is necessary to work with a local K3  with a  singularity  $A_1$. The deformed geometry is   given  by the product of the complex $ {\bf C}$ plane and a two sphere $S^2$.   Since $\Pi_q(X \times Y)=\Pi_q(X )\times \Pi_q(Y)$,  we have the following remarkable relation 
 \begin{equation}
 \pi_4(A_1)\sim\pi_4(S^2)=Z_2.
\end{equation}
 The two charges of the Yang monopole are related to the  two ways the geometry allows a D$4$-brane to wrap on  it.

 Another possible interpretation takes into account the fact that a locally  K3 surface (\ref{A1}) has a  $Z_2$  symmetry  acting as follows
\begin{equation}
x \to w x,\qquad
y \to {\bar w}y,\qquad
z \to w z,
\end{equation}
where $w^2=1$.   Wrapping D4-branes over  such a geometry 
gives two  configurations. Each one corresponds to an equivalent class of the $Z_2$ group, which   can be identified with 
 the center of the $SU(2)$  gauge   symmetry in six dimensions.  The wrapped  D4-brane  over  the K3 surface is 
sensitive  to this $Z_2$ symmetry  and,  thus,    it carries $\{+1,-1\}$  charges. A general value of the charge, which corresponds to a non spherical solution, can be obtained by wrapping an arbitrary number of D4-branes on K3. These solutions are the multinstanton configuration on the $S^4$.

It is known that the energy of the Yang monopole diverges linearly in space-time. This fact is not manifest in our geometric construction. We believe that this property can be understood in the effective field theory description.

\section{Relation with the heterotic M-theory Yang monopole configuration}
 As explained in the  introduction,  the authors of \cite{BGT} have  suggested a  Yang monopole  representation  
with two $SU(2)$ gauge  factors obtained by breaking the  $E_8 \times  E_8$  heterotic gauge symmetry in ten
 dimensions. This breaking, $E_8 \to E_7\times SU(2)$, could be related with the fact that the extremes of the M5-branes
 are located on the M9-branes and they  are just the core of the Yang monopole.  On the Type IIA side however   there is only one $SU(2)$ 
factor  which comes from  a D2-brane  wrapped 
around the collapsing $S^2$  inside the K3 surface. Lifting to  M-theory the nature of this difference is apreciated. 
Reduction  from 11 to 6 dimensions with sexteen supercharges  can be performed  in two dual  ways depending on the action
 of the $Z_2$ symmetry on the five dimensional internal space $S^1\times T^4$.  In  the heterotic realization
 of M-theory, the $ Z_2$ symmetry acts on the  $S^1$ factor giving rise the segment between the two M9-branes,  
while in   the  type IIA, M-theory,   the symmetry acts on  the $T^4$ factor producing the $K3$ geometry. However,  since  these two M-theory compactifications
 are dual in six dimensions, 
  the  two above string  Yang  monopole realizations
 should be connected.  In what follows,  we will argue on how they could  be related. 

The K3 surface has two  possible  constructions as the target space of a sigma model. They  depend on the R-symmetry
 of the supercharges. Previously, we have mainly concerned with   $N=2$ sigma model, where the R-symmetry
 is supported by a $U(1)$ group, and the $K3$ target space  gets  manifested as a Khaler manifold. Now,  let  us   use the
 other realization of $K3$ where the manifold is hyperkhaler and the corresponding sigma model involves
 eight supercharges and  has a $SU(2)$ R-symmetry.
 Then, the  $\{+1,-1\}$  charges of the Yang monopole  can  be explained by 
 physical arguments  when the K3 surface is constructed in terms of   $N = 4$ sigma model.   This 
  is  related to the   heterotic M-theory configuration where  the Yang monopole has two  copies in the boundaries
  of the  M5-brane suspended between  two  M9-branes.
Theses copies, with charges $+1$ and $-1$,   might  be understood as  two hypermultiplets  appearing  in the hyperkhaler quotient
construction of the $A_1$ local manifestation of $K3$.

The six dimensional $SU(2)$   Yang Mills   theory can  also be  obtained from  a K3 surface 
that is realized in terms of $N = 4$  supersymmetric sigma model. This sigma model has only one  $ U(1)$  gauge group,   two  hypermultiplets   with 
 charges  $(q_1,q_2)$, and  one isotriplet FI coupling ${\vec \xi} = (\xi_1, \xi_2, \xi_3)$ \cite{AW,B}. The sigma model gauge symmetry is related  to  the Cartan subgroup of the six dimensional gauge group.  In this construction,
the K3 surface is expressed by the vanishing condition of the  following 
D-terms
\begin{equation}
\sum_{i=1}^2q_i( \phi_i^\alpha {\bar\phi}_{i\beta}+\phi_{i\beta}{\bar\phi}_i^\alpha)-{\vec \xi}{\vec \sigma}^{\alpha}_{\beta}=0. \label{hyper}
\end{equation}
The double index $(i, \alpha)$  of the scalars refers to  the component field doublets ($\alpha$) of the  two  hypermultiplets ($i$), and
$\sigma$ are the traceless $2 \times 2$ Pauli matrices. The condition under which the gauge theory flows in the infrared to 2d
$N = 4$  superconformal field theory, which   is also the condition to have a  hyperkhaler 
Calabi-Yau background, is 
\begin{equation}
q_1+q_2=0.
\end{equation}
This equation has different solutions that can be seen as redefinitions of the coupling constant $\vec \xi$. 
Due to its conformal invariance, the theory does not get affected by redefinitions of $\vec \xi$, so the charge can be fixed to $-1$ and $+1$. 

Let us discuss the construction of $K3$ in this case. The starting point consists of two hypermultiplets with four scalars each. They can be expressed as ${\bf R^4}\times {\bf R^4}$.
The gauge invariance of each hypermultiplet (with $+1$ and $-1$ charge) under the $U(1)$ symmetry, together with the invariance under the $SU(2)$ R-symmetry that rotates the supercharges, enables us to express the $K3$ locally as the following homogenous space
\begin{equation}
\frac{{\bf R}^4 \times {\bf R}^4}  {U(1)\times SU(2) }.
\end{equation}
There is a $Z_2$ symmetry that interchanges the two hypermultiplets (the two ${\bf R}^4$ factors). We interpret the two ${\bf R}^4$ factors with their corresponding charges  as the two Yang monopole copies which are the boundaries of the M5-branes  on the two M9-branes. 
At this stage, we can ask the following question,
which is the role of the $SU(2)$ R-symmetry in this construction?
The answer of this question lies on the association of the R-symmetry with the instantonic nature of the M5-brane in the context of the heterotic M-theory picture.

\section{Discusion}
In this  paper, a  Type IIA geometric  realization of the  Yang monopole in  six dimensions is given.  For this purpose, it has been   used  the result of the duality  between 
 Type IIA superstring    compactified on    the   K3 surface  and    heterotic superstring  on  $T^4$.   The $SU(2)$ gauge symmetry of the Yang monopole
 has been considered as the  enhanced gauge symmetry  corresponding to  shrinking 2-cycles inside  the  K3 surface.   
We have shown  that  the  Yang monopole  comes up by wrapping D-branes on the K3 non-trivial
 cycles. In this way, the properties of the Yang monopole  have been  encoded in the K3 surface features. 
 Then we  have analyzed how this realization    can be  related  to the result of \cite{BGT} in  which a  heterotic M-theory picture is involved.

The Yang monopole interpretation given in this paper  opens  some  interesting questions in the  connection with
 the  type IIA superstring black hole physics in six dimensions. In this case, the   black hole, 
  the black string and the  black  membrane appear  in the same footing. 
 This is a consequence of the fact  that the  physical dimensions of two dual configurations
 are related by $p+q=D-4=2$, where $p$ and $q$ are the spatial  dimensions of the branes
 and $D=6$ is  the dimension of the space-time where they live.  It should be interesting to connect our
 D-brane Yang monopole construction with the six dimensional black solution of Einstein equations.  This will be reported elsewhere.

When the gravitational interaction is taken into account,  the Yang monopole curves the space-time   which develops
event horizon. The geometry is similar to the one of the Reiner-Nostrom black hole, and 
 it has been  also   obtained in the presence of a  cosmological constant\cite{GT}. In \cite{DS}, 
 this geometry  has been analyzed for the Nariai type solutions when the cosmological and black hole
 horizons coalesce. Extremal geometries can be candidates for BPS type solutions where the attractor mechanism could apply.
We hope the results of this paper can be used to address this point.

\section*{Acknowledgements} \nonumber
We thank   M. Asorey, L. J. Boya, E. H. Saidi and  P.  K. Townsend for  discussions.  This work has been supported by MCYT ( Spain) under grant FPA
2003-02948.

\end{document}